\documentclass[a4paper, amsfonts, amssymb, amsmath, reprint, showkeys, nofootinbib, twoside]{revtex4-1}
\usepackage[english]{babel}
\usepackage[utf8]{inputenc}
\usepackage[colorinlistoftodos, color=green!40, prependcaption]{todonotes}
\usepackage{amsthm}
\usepackage{xcolor}
\usepackage{graphicx}
\usepackage[T1]{fontenc}
\usepackage[pdftex, pdftitle={Article}, pdfauthor={Author}]{hyperref} 
\begin{document}
\title{Holography and matter creation revisited}

\author{V\'ictor H. C\'ardenas$^1$}
\email{victor.cardenas@uv.cl}

\author{Miguel Cruz$^2$}
\email{miguelcruz02@uv.mx}

\author{Samuel Lepe$^3$}
\email{samuel.lepe@pucv.cl}

\affiliation{$^1$Instituto de F\'{\i}sica y Astronom\'ia, Universidad de Valpara\'iso, Gran Breta\~na 1111, Valpara\'iso, Chile \\
$^2$Facultad de F\'{\i}sica, Universidad Veracruzana 91000, Xalapa, Veracruz, M\'exico \\
$^3$Instituto de F\'{\i}sica, Facultad de Ciencias, Pontificia Universidad Cat\'olica de Valpara\'\i so, Avenida Brasil 2950, Valpara\'iso, Chile 
}

\date{\today} 

\begin{abstract}
The consideration of the holographic dark energy approach and matter creation effects in a single cosmological model is carried out in this work accordingly, we discuss some cases of interest. We test this cosmological proposal against recent observations. Considering the best fit values obtained for the free cosmological parameters, the model exhibits a transition from a decelerated cosmic expansion to an accelerated one. The deceleration-acceleration transition can not be determined consistently from the coincidence parameter for this kind of model. At effective level the parameter state associated to the whole energy density contribution describes a quintessence scenario in the fluid analogy at present time.       
\end{abstract}

\keywords{holography, cosmic expansion, dark energy, creation}

\maketitle

\section{Introduction}

Our most simple explanation for the observed cosmic acceleration \cite{obs} generally requires of the dark energy concept, a strange component as its own name. Its complex nature it is not totally unveiled yet but nowadays, several theoretical approaches result to be useful to reproduce in good manner its effects at large scales. In Ref. \cite{entropic} the idea of a fundamental origin for the accelerated expansion is motivated by the consideration of non-equilibrium thermodynamics in the gravity description, in other words, the universe evolves in a non-adiabatic way. The contribution from non-equilibrium phenomena lead to a back-reaction on the gravitational field equations with observational consequences: cosmic acceleration ensues from entropy production. A non-adiabatic universe seems to be more interesting and consistent than the one predicted by the cosmological standard model usually known as $\Lambda$CDM. In this model the cosmic evolution is adiabatic and leads to a null temperature for dark energy when the fluid analogy is used. This result looks quite obvious since dark energy is modeled by a constant (omnipresent) term known as cosmological constant. However, some other implications emerging from $\Lambda$CDM at thermodynamics level are physically meaningless, see for instance \cite{grandon}.\\

Non-adiabatic cosmic evolution is not a new idea, it is well known that for perfect fluids the transfer of heat is not possible due to the absence of entropy production; in order to explain successfully some of the astrophysical processes found in our universe, we must ensure such heat transfer in our description. A variable entropy scenario motivated the examination of dissipative losses in the fluid. However, the dissipative scheme applies only in the near equilibrium condition and its extension beyond near equilibrium processes implies facing great complexity, for an introductory lecture on the topic see \cite{viscous} and other works that can be found in the literature. Fortunately, other (and more simple) cosmological scenarios allow the production of entropy. We focus on the matter production case\footnote{We must have in mind that this kind of process is based on fundamental quantum aspects of spacetime \cite{birrel}.}; the key idea in this framework relies on the use of a balance equation\footnote{The balance equation for the particle number density has the form, $\dot{n}+3nH = n\Gamma$, in a flat Friedmann-Lemaitre-Robertson-Walker (FLRW) spacetime. In this case $\Gamma$ must be considered as the particle source (sink).} for the created particles together with the Einstein equations, when this is properly established and combined with the second law of thermodynamics, the stress tensor changes by an additional contribution which is usually interpreted as a pressure term and generally denoted as $p_{c}$, in turn the aforementioned pressure depends on the production rate, $\Gamma$. In this kind of model the entropy production comes from two mechanisms\footnote{The entropy density can be written as, $s^{\mu}_{;\mu}=ns(\dot{N}/N) + n\dot{s}$, where $N$ is the total number of particles given in a volume $a^{3}$, i.e., $N=na^{3}$. The adiabatic case implies $\dot{s}=0$.}. The first due to the growth of the number of fluid particles, and the second is associated to the increase in entropy per fluid particle, generally this latter amount of entropy is induced by dissipative processes and/or shear stress. Then we refer to adiabatic particle production a process in which the entropy per fluid particle is constant or in other words, the fluid description does not take into account dissipative effects. In this case the entropy production comes entirely from the proliferation of particles in the fluid. The assumption of constant entropy per fluid particle is equivalent to assume that created particles admit a perfect fluid description rapidly after their creation. We refer the reader to Ref. \cite{adiabatic}, where some relevant works on adiabatic matter production can be found.\\ 

The congruous formalism for matter creation processes in the context of FLRW geometries was initially proposed in \cite{prigogine}. The contemporary description of the gravitational field has been possible due to the understanding of the junction between thermodynamics and general relativity. The works of Bekenstein and Hawking \cite{bhthe} are the cornerstone of the black holes thermodynamics; due to their contribution we can associate temperature and entropy to this kind of objects. These seeds found fertile field in cosmology quickly; Bekenstein himself proposed a cosmological extension of his entropy bound\footnote{The upper bound on the entropy proposed by Bekenstein is of the form: $S_{IS} \leq S_{B} = 2\pi E R$, where $E$ is the energy of the system and $R$ its size \cite{beken1}.} for any isolated physical systems by setting up the particle horizon as the size of the system \cite{beken2}. In general terms, the bound on the entropy proposed by Bekenstein implies the holographic principle. The formal extension of the holographic principle to the cosmological context can be found in \cite{Fischler:1998st}. Even now the holographic principle is seen as a breakthrough formalism that has changed radically our understanding about the gravitational field \cite{witten}. In this sense, we find interesting the proposal given in Refs. \cite{sing, singb} in order to describe the dark energy enigma since both approaches matter production effects and holography are well founded from the thermodynamics point of view. Some works about thermodynamics in these topics can be found in \cite{ther1, ther2, ther3}. Additionally, flat cosmological models in which matter creation effects are considered without any other contribution lead to condition, $\Omega_{m,0} =1$, at present time; being $\Omega_m$ the fractional energy density parameter associated to the matter sector; nowadays the value for $\Omega_{m,0}$ is estimated around $0.3$. Therefore we must consider an extra component, such that when added with the matter sector gives, $\Omega_{\mathrm{total}}=1$, this extra contribution can come from dark energy. The aim of this work is to provide a clear discussion of the emergent cosmological scenario coming from the combination of holography and matter creation effects supported by observational data.\\

The organization of this work is the following: in section \ref{sec:hde} we provide the main elements of the cosmological model. We discuss briefly some aspects of the holographic approach and the way is combined with matter creation effects. We also comment about a simple model considered earlier in other works and we extend it to our proposal by the introduction of the Barrow exponent in the holographic cutoff and a variable production rate. In section \ref{sec:data} we perform the statistical analysis of both models in order to compare them and we also include three special cases of interest. Section \ref{sec:final} is devoted to the final comments of our work. The units chosen for this work are $8\pi G=c=k_{B}=1$.   

\section{Holographic approach and matter creation: the way these ideas are combined}
\label{sec:hde}

A seminal work on holography is due to t'Hooft \cite{tHooft:1993dmi}. In this work t'Hooft proposed that a proper physical description in a region of the universe only needs the degrees of freedom of the {\it surface} of that region, and not the ones in the bulk, i.e., as an hologram. This requires the entropy of a region must not exceed its area in Planck units.

As mentioned before, the first proposal which incorporates holography into cosmology was presented by Fischler and Susskind in \cite{Fischler:1998st}. After this, a large number of works have explored different implementations of this idea (see a review on the topic in \cite{Bousso:2002ju}). A very interesting proposal is given by Cohen et.~al in Ref. \cite{cohen}. They proposed an upper limit for the energy density based on the mass of a black hole of the same size can form. In this case the dark energy density evolves as
\begin{equation}\label{eq:roxL}
    \rho_x = 3b^2L^{-2},
\end{equation}
where $b$ is a constant in the range $(0,1)$ and the characteristic length $L$ was identified with the Hubble size $L=H^{-1}$. The interesting fact about the cutoff (\ref{eq:roxL}) was that by assuming $b$ being of the order one, the energy density acquires a value quite close to the current dark energy contribution. However, Hsu \cite{Hsu:2004ri} demonstrated that although the value of the energy density is correct, the EoS parameter does not, the assumption (\ref{eq:roxL}) leads to the case $\omega=p/\rho=0$. So this model cannot be considered as a viable candidate for dark energy.

A self consistent holographic dark energy model was proposed by Li \cite{li}, in which the dark energy density evolves as (\ref{eq:roxL}) with the characteristic length $L$ being the future event horizon defined as
\begin{equation}
    L = a\int_{t}^{\infty} \frac{dt}{a}.
\end{equation}
Because the length depends on the evolution of the model, the test of this model against data requires to solve a coupled system in terms of the redshift $z$ (see \cite{Li:2013dha})
\begin{eqnarray}
    \frac{E'}{E} = -\frac{\Omega_x}{1+z} \left(\frac{\sqrt{\Omega_x}}{c}+\frac{1}{2} - \frac{\Omega_r +3}{2\Omega_x} \right), \\
    \Omega_x' = -\frac{2\Omega_x(1-\Omega_x)}{1+z} \left( \frac{\sqrt{\Omega_x}}{c}+\frac{1}{2} + \frac{\Omega_r}{2(1-\Omega_x)} \right),
\end{eqnarray}
where $E(z):=H(z)/H_0$ is the normalized Hubble function, $E':=dE/dz$, $\Omega_x$ is the fractional energy density parameter associated to the dynamical dark energy and $\Omega_r$ is the radiation density parameter. The result of this test indicates a preferred value of $c = 0.508 \pm 0.207$.

Instead using a different infrared cutoff as Li did, we can stick to the original proposal of Cohen et al., \cite{cohen} with $L=H^{-1}$ and use the matter creation process as the source for cosmic acceleration. This was the idea of Kumar and Singh developed in a coupled of recent papers \cite{sing} and \cite{singb}. Following the line of reasoning of Ref. \cite{sing}, the authors considered a model comprised by an holographic dark energy ($x$) component together with a matter ($m$) component coming from a matter creation process. The equations that describe the system in a spatially flat FLRW background are
\begin{align}
& 3H^2 = \rho_m + \rho_x, \label{eq:fried1}\\
& 2\dot{H} + 3H^2 = -p_x-p_m-p_c, \label{eq:accel}\\
& p_c = -\frac{(\rho_m + p_m)}{3H}\Gamma, \label{eq:creatpre}\\
& \dot{\rho}_m + 3H (\rho_m + p_{m}) = \Gamma \rho_m, \label{eq:mat} \\
& \dot{\rho}_x + 3H (\rho_x + p_x) = 0, \label{eq:fried2}
\end{align}
where $\rho_i, p_i$ are the energy density and pressure for the i-th constituent being $i=x,m$ for dark energy and dark matter respectively. Here $p_{c}$ accounts for the pressure from matter creation and $\Gamma$ is the particle creation rate. The expression for matter creation pressure it is not imposed by hand, emerges from the consideration of constant specific entropy (per particle), i.e., an adiabatic process, as stated in reference \cite{pdu1}, in which the covariant formulation of the Prigogine results found in \cite{prigogine} was proposed. Also notice that $p_{c}$ contributes negatively. Although interesting, the analysis of this model in particular needs a revision and correction in some aspects that we would like to highlight below.

\subsection{A simple model}
\label{sec:simple}

Let us start discussing the simplest model for matter creation process. The one considering a constant particle rate given as $\Gamma = 3\beta H_0$ and let us assume the holographic dark energy model of Cohen $\rho_x = 3b^2H^2$. We also consider a dust EoS for dark matter, $p_m=0$. This model was proposed in Ref. \cite{sing}. From Eq. (\ref{eq:accel}) one gets
\begin{equation}\label{eq:pxsimp}
    p_x = -\rho_m-\rho_x -2\dot{H} + \beta \frac{H_0}{H}\rho_m.
\end{equation}
Now, inserting this pressure in Eq. (\ref{eq:fried2}) we obtain 
\begin{equation}\label{eq:heqsimp}
    \frac{\dot{H}}{H} + \frac{3}{2}\left(H-\beta H_0 \right)=0,
\end{equation}
which is not the equation quoted and solved in Ref.\cite{sing}. The main reason for the discrepancy is the assumption made in \cite{sing} of a constant EoS for the dark energy component, $p_x=\omega_x \rho_x$ with $\omega_x = \mbox{constant}$. Clearly this is not possible to assume, because of (\ref{eq:pxsimp}). Writing Eq. (\ref{eq:heqsimp}) in terms of the normalized Hubble function and as a function of redshift we get
\begin{equation}
    \frac{dE}{dz} - \frac{3(E-\beta)}{2(1+z)}=0,
    \label{eq:pdu1}
\end{equation}
whose solution is
\begin{equation}\label{eq:edzsimp}
    E(z) = \beta + (1-\beta)(1+z)^{3/2}. 
\end{equation}
Due to the proportionality of the holographic dark energy to $H^{2}$, we obtain that the expression for $E(z)$ given above is independent of the parameter $b^2$ that controls the dark energy density and so it is not important to fit the observational data. We would like to comment that for this holographic approach, Eq. (\ref{eq:heqsimp}) and in consequence (\ref{eq:pdu1}), resemble the matter creation model studied in Ref. \cite{pdu2}, which connects both accelerated phases of the universe with some reliability. This similarity is also valid for the general model explored below with $\gamma=0$. 

\begin{figure}[h!]
    \centering
    \includegraphics[width=8cm]{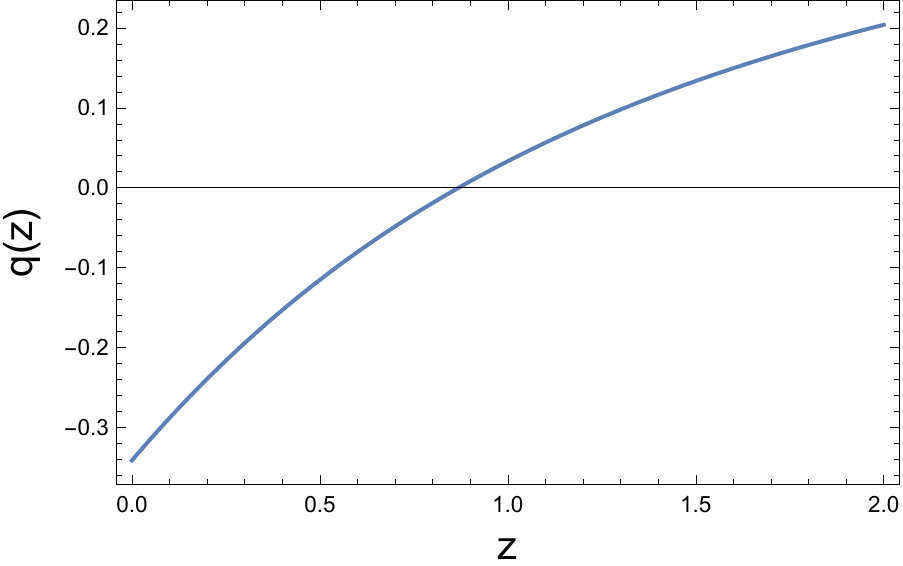}
    \caption{Deceleration parameter from Eq.(\ref{eq:qdzsimp}) using $E(z)$ from (\ref{eq:edzsimp}). Notice the transition from decelerated to an accelerated phase. Here we use $\beta=0.561$ based on the statistical analysis described in the text. }
    \label{fig:qdzsimp}
\end{figure}

Notice that this model has a non trivial deceleration parameter. In fact, from (\ref{eq:heqsimp}) we get
\begin{equation}\label{eq:qdzsimp}
    q = \frac{1}{2} - \frac{3}{2}\beta \frac{H_0}{H},
\end{equation}
which after using the explicit expression for $E(z)$ from (\ref{eq:edzsimp}) we obtain the curve shown in Fig.(\ref{fig:qdzsimp}).
Clearly, this model is able to describe the transition from a decelerated to an accelerated universe with a single free parameter $\beta$, the parameter that mediates the rate of particle creation. If $\beta=0$ from (\ref{eq:edzsimp}) we obtain the usual Einstein-de Sitter solution with $q=0.5$ from (\ref{eq:qdzsimp}).

Clearly the analysis of a model that mixes matter creation and holographic dark energy should be redone. In the next subsection we describe a more general model for both matter creation and the holographic dark energy. In section \ref{sec:data} we perform a statistical analysis of these models using observational data.

\subsection{Exploring a general model}

In this section we will explore a generalization of the model discussed above. For the holographic model we will consider now the Barrow holographic cutoff and the Hubble scale \cite{barrow}, given as
\begin{equation}
    \rho_{x} = 3\gamma H^{2}_{0}E^{2-\Delta}, \label{eq:roxhol1}
\end{equation}
where $\Delta$ is a constant parameter. Note that in the case $\Delta = 0$, we recover in the previous expression the holographic cut proposed by Cohen \cite{cohen}. If we perform a comparison between the above expression and the usual holographic formula, we observe that our expression depends explicitly on the physical parameter $H_{0}$, this is due to units consistency. Then, $\gamma$ is a dimensionless constant parameter that must lie in the interval, $0 < \gamma < 1$. In our description we will take into account a more general expression for the matter production rate provided in Ref. \cite{waga}
\begin{equation}
    \Gamma = 3\beta H_0 \left( \frac{H}{H_0}\right)^{\alpha}.
    \label{eq:waga}
\end{equation}
It is worthy to mention that for $\alpha = 0$, we obtain a constant production rate, as discussed in the previous section. For a general expression of $\Gamma$, the Eq. (\ref{eq:pxsimp}) takes the form
\begin{equation}
    p_{x} = -\rho_{m} -\rho_{x} - 2\dot{H}+\rho_{m}\frac{\Gamma}{3H}.
    \label{eq:depressure}
\end{equation}
On the other hand, inserting the above expression in the continuity equation (\ref{eq:fried2}) and considering the Eqs. (\ref{eq:roxhol1}) and (\ref{eq:waga}), we can obtain the following expression in terms of the normalized Hubble parameter  
\begin{equation}
    \frac{\dot{E}}{E} + \frac{3\left(1-\gamma E^{-\Delta} \right)\left(H_{0}E-\beta H_{0}E^{\alpha}\right)}{2\left(1-\gamma E^{-\Delta} \right) + \Delta \gamma E^{-\Delta}} = 0.
\end{equation}
Note that for the case $\alpha = \Delta = 0$ we recover the Eq. (\ref{eq:heqsimp}). Alternatively, we can write the evolution equation for the normalized Hubble parameter as a function of the redshift from the previous result, yielding
\begin{equation}
    \frac{dE}{dz} - \frac{3\left(1-\gamma E^{-\Delta} \right)\left(E - \beta E^{\alpha}\right)}{(1+z)\left[2\left(1-\gamma E^{-\Delta} \right) + \Delta \gamma E^{-\Delta}\right]}= 0, \label{eq:turning2}
\end{equation}
The dependence on the redshift of this differential equation together with the general case given by the Barrow holographic dark energy and the $\alpha$ exponent for the matter production rate, lead us to a numerical treatment. Additionally, the differential equation (\ref{eq:turning2}) does not coincide with the one studied in \cite{sing} when $\Delta = \alpha = 0$. The above expression will be useful to constraint the cosmological parameters of the model. Some comments are in order regarding the previous equation. We can have turning points in $E(z)$ characterized by $dE(z)/dz =0$, for the following cases: $\gamma = 1$ and $\Delta = 0$, $\beta = \alpha = 1$ and $\Delta = \pm \infty$. The existence of turning points in holographic dark energy models was extensively discussed in Ref. \cite{hdetens}, they play a crucial role on the alleviation of the $H_{0}$ tension. However, their presence imply the acceptance of a running value for $H_{0}$, i.e, the breakdown of homogeneity and isotropy of the universe; which are the principal assumptions in the FLRW cosmology. As we will see below in the statistical analysis, the existence of turning points in our model is kicked out according to the constrained values for the set of cosmological parameters. On the other hand, taking into account Eqs. (\ref{eq:roxhol1}) and (\ref{eq:depressure}), we can construct an explicit form for the parameter state, $\omega$, if we restrict ourselves to a barotropic EoS
\begin{equation}
    \omega_{x}(z) = -1 + \frac{(2-\Delta)(1 - \gamma E^{-\Delta})(1-\beta E^{\alpha-1})}{2-\gamma(2-\Delta)E^{-\Delta}}.
    \label{eq:omegagen}
\end{equation}
As can be seen, the parameter state for the dark energy fluid depends on the matter production rate and deviates from the $\Lambda$CDM model value since evolves as a function of the redshift; as long as a numerical solution for $E(z)$ is obtained from Eq. (\ref{eq:turning2}), we can infer the behavior for $\omega(z)$. Additionally, if we combine the Eqs. (\ref{eq:mat}) and (\ref{eq:fried2}) we can identify the parameter state associated to the total energy density, $\rho_{T}=\rho_m+\rho_x$, coming from the holographic dark energy and matter creation effects; we again consider a barotropic EoS for $\rho_{T}$ and $p_{T}$, then the continuity equation for the total energy density takes the usual form $\dot{\rho}_{T}+3H(1+\omega_{\mathrm{eff}})\rho_{T} = 0$, where the effective parameter state is given explicitly as follows
\begin{eqnarray}
    \omega_{\mathrm{eff}}(z) &=& \frac{\omega_{x}(z)+\frac{\rho_{m}}{\rho_{x}}\frac{\Gamma}{3H_{0}E}}{1+\frac{\rho_{m}}{\rho_{x}}}, \nonumber \\
    &=& \frac{\gamma \omega_{x}(z) + \beta (1-\gamma E^{-\Delta})E^{\alpha + \Delta - 1}}{\gamma + (1-\gamma E^{-\Delta})E^{\Delta}}.
    \label{eq:omegaeff}
\end{eqnarray}
In order to explore the cosmological results of the model, in the following section we will discuss the behavior of the latter expression based on the statistical analysis results.

\section{Cosmological parameters}
\label{sec:data}

In this section we study the performance of the models to fit the observational data and characterize the free parameters constrained by the observations. Here we use 
cosmic chronometers by using a sample of measurements of $H(z)$. 
The $H(z)$ data consists in 57 data points, $31$ of which were obtained by cosmic chronometers \cite{Moresco:2016mzx,Moresco:2015cya,2014RAA....14.1221Z,2010JCAP...02..008S,2012JCAP...08..006M,Ratsimbazafy:2017vga} and 26 data points obtained from BAO data \cite{BOSS:2014hwf,2012MNRAS.425..405B,Chuang:2013hya,BOSS:2013igd, Bautista:2017zgn,Gaz34,Oka37,Wang33,Chuang28,Alam38,Anderson32,Busca36}.
The analysis were performed using the code EMCEE \cite{2013PASP..125..306F} which consist in a Python implementation of the affine-invariant ensemble sampler for Markov chain Monte Carlo (MCMC) proposed by Goodman and Weare \cite{GW2010}. 

\subsection{The analysis of the simple model}

Let us start with the simple model described in section \ref{sec:simple}. Notice that although the model is characterized by two parameters, $\beta$ for the matter creation process and $\gamma$ for the holographic dark energy model, the Hubble function is only dependent of $\beta$. In some sense, $\beta$ in (\ref{eq:edzsimp}) acts as $\Omega_{\Lambda}$ in the $\Lambda$CDM model.

\begin{figure}[h!]
    \centering
    \includegraphics[width=7cm]{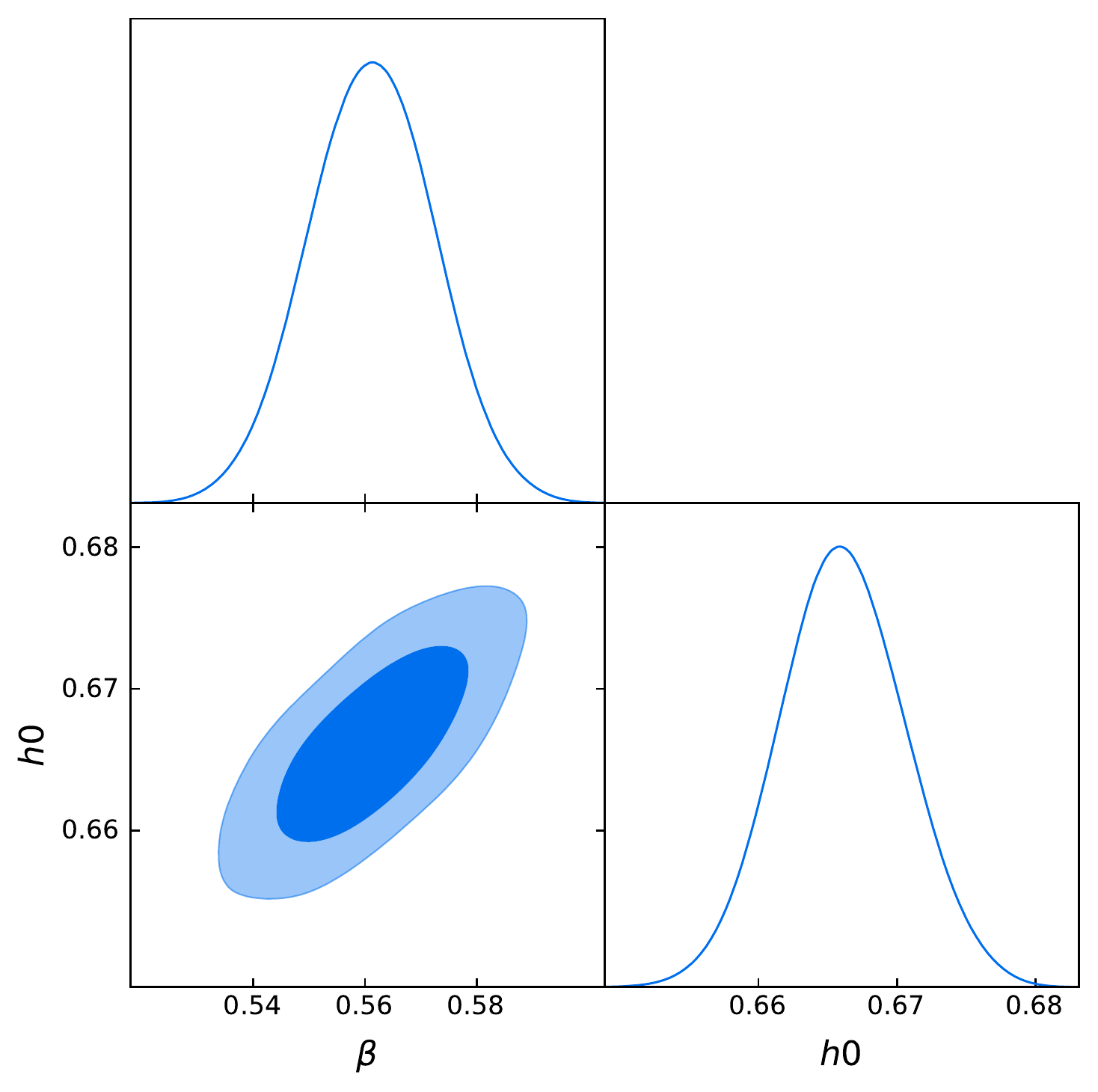}
    \caption{We plot the posteriors for the parameters $(\beta, h)$ of the model defined by (\ref{eq:edzsimp}) using the $H(z)$ data.}
    \label{fig:postsimp}
\end{figure}

Let us use the observational data, measurements of the Hubble function $H(z)$ mentioned before, to constraint the free parameters. In this simple case we use the chi square defined by
\begin{equation}\label{eq:chi2hz}
    \chi^2_{H(z)} = \sum_{i=1}^{57} \left(\frac{100hE(z_i,\theta) - H(z_i)}{\sigma_{H_i}}\right)^2,
\end{equation}
where we put in tension the data $H(z_i)$ with its uncertainty $\sigma_{H_i}$ against the theoretical prediction $E(z_i,\theta)$. In this simple model, the free parameters are only two, $\theta =(h, \beta)$, because of (\ref{eq:edzsimp}). The best fit values obtained after 3000 chains were: $\beta = 0.561 \pm 0.011$ and $h = 0.6660\substack{+0.0046\\-0.00448 }$. The 1D and 2D posteriors are shown in Fig.(\ref{fig:postsimp}).

\subsection{Analysis of the full model}

In this case we use the chi square (\ref{eq:chi2hz}) with the $H(z)$ data to constraint $E(z)$ given by the numerical solution of Eq. (\ref{eq:turning2}), where $\theta$ is the vector parameters given by $\theta = (\alpha, \beta, \gamma, \delta)$. The statistical analysis then considers five free parameters, $(h,\alpha, \beta, \gamma, \delta)$.
 
\begin{figure}[h!]
    \centering
    \includegraphics[width=8cm]{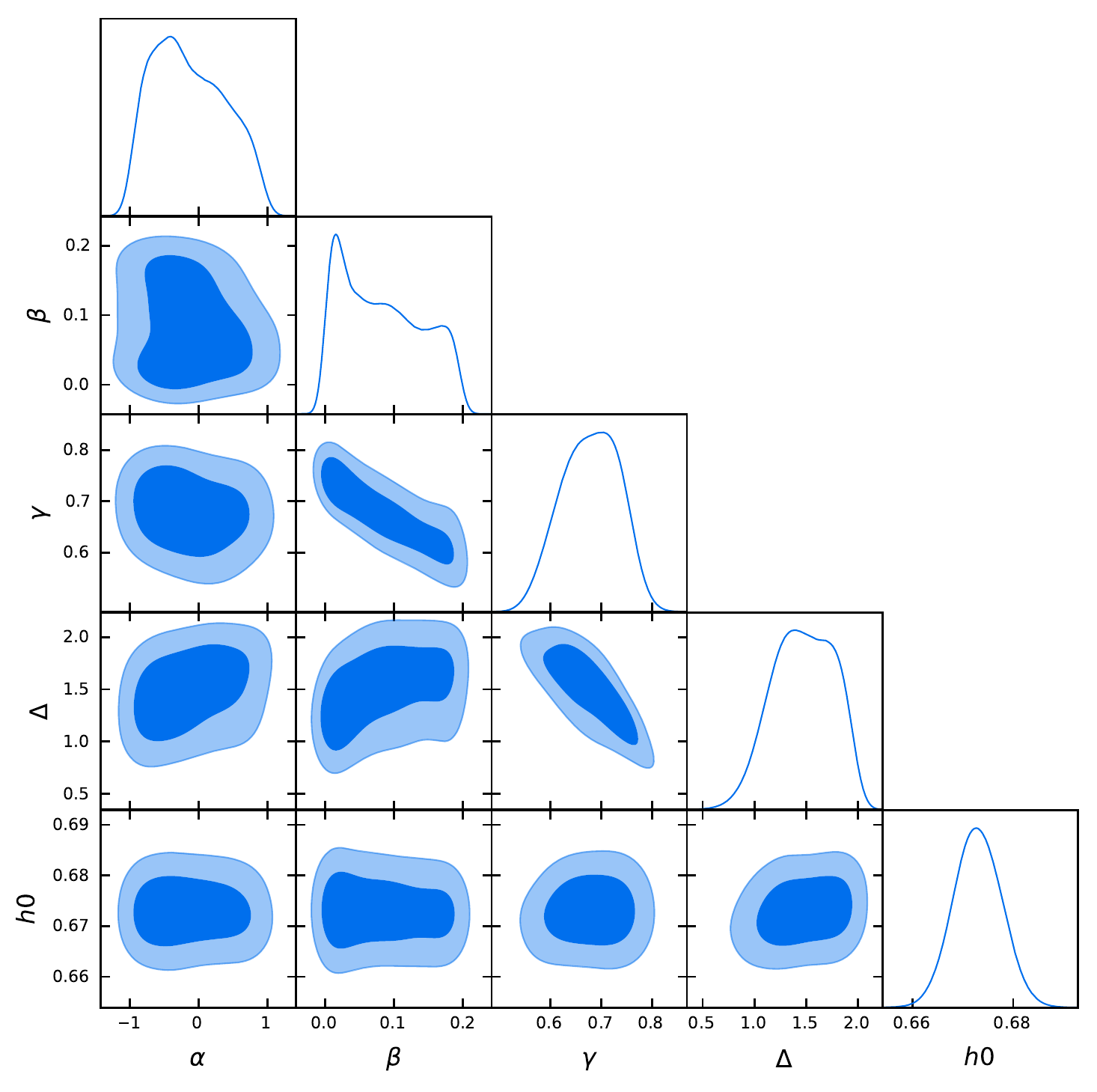}
    \caption{We plot the posteriors for all the parameters using the $H(z)$ data.}
    \label{fig:contours}
\end{figure}
We consider uniform priors on $\gamma>0$ and $\beta>0$ because both have to be positive, the first by the strong energy condition and the second by the second law of thermodynamics. We have also added a prior in the Hubble constant $H_0$ from Planck, $H_0=0.674 \pm 0.005$.

The result of the statistical analysis is given in the first column of Table (\ref{tab:results}) with a $\chi^2_{min} = 29.7$.
\begin{table}
\begin{ruledtabular}
\begin{tabular}{lcccc}
Params & Full model & case (i) &  case (ii) & case (iii) \\
$\alpha$ & $-0.19 \substack{+0.67 \\ -0.52}$ & $0$ & $-1$ & $-0.55 \substack{+0.17 \\ -0.16}$  \\
$\beta$ & $0.08\substack{+0.08 \\ -0.06}$ \ & $0.628 \substack{+0.028 \\ -0.033}$ & $0.013 \substack{+0.06 \\ -0.09}$ & $0.642 \substack{+0.026 \\ -0.025}$ \\
$\gamma$ & $0.681\substack{+0.059\\-0.062}$ & $0.27\substack{+0.28\\-0.17}$ & $0.72 \substack{+0.12 \\ -0.08}$ & -\\
$\Delta$ & $1.49\substack{+0.33\\-0.32}$ & $-0.29\substack{+0.20\\-0.39}$ & $1.40 \substack{+0.45 \\ -0.65}$ & $0$\\
$h$ & $0.673\substack{+0.005\\-0.005}$ & $0.670\substack{+0.005\\-0.004}$ & $0.673 \substack{+0.005 \\ -0.005}$ & $0.673 \substack{+0.004 \\ -0.005}$\\
\end{tabular}
\end{ruledtabular}
\caption{\label{tab:results}Constrained values for the free parameters of each cosmological model.}
\end{table}
It is important to notice that $\beta$ takes a small value consistent with zero indicating that matter production effects are not viable to describe the cosmic acceleration, then in this case the Barrow holographic dark energy is managing the late times cosmic evolution. A similar situation occurred in Ref. \cite{waga}, where the dark energy is characterized by the cosmological constant. Note that in our approach the holographic model becomes viable to describe the accelerated expansion since deviates from the Cohen model due to the introduction of the exponent $\Delta$, this assertion can be seen by setting $\beta=0$ in Eqs. (\ref{eq:omegagen}) and (\ref{eq:omegaeff}). A relevant result in this analysis is the positivity of the parameter $\beta$, which ensures $\Gamma > 0$ leading to $p_{c} < 0$.

\begin{figure}[h!]
    \centering
    \includegraphics[width=8cm]{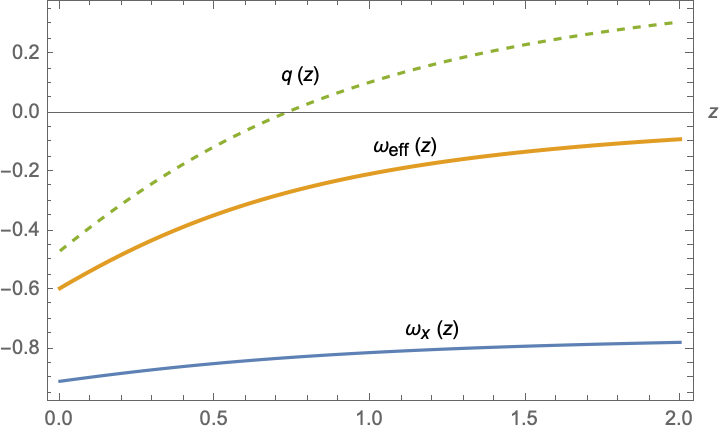}
    \caption{Cosmological parameters $q(z)$, $\omega_{x}(z)$ and $\omega_{\mathrm{eff}}(z)$ of the generalized model.}
    \label{fig:genparams}
\end{figure}

In Fig. (\ref{fig:genparams}) we show the behavior of the parameter state given in Eqs. (\ref{eq:omegagen}) and (\ref{eq:omegaeff}) using the best fit values obtained from the statistical analysis for the generalized model. We also evaluate the deceleration parameter, $q$, which can be obtained from Eq. (\ref{eq:turning2}) as follows
\begin{eqnarray}
    q &=& - 1 - \frac{\dot{H}}{H^{2}} = - 1 - \frac{\dot{E}}{H_{0}E^{2}} = -1 + \frac{(1+z)}{E}\frac{dE}{dz},\nonumber \\
    &=& -1 + \frac{3}{2}\frac{\left(1-\gamma E^{-\Delta} \right)\left(1-\beta E^{\alpha-1}\right)}{\left[1-\left(\gamma/2 \right)(2-\Delta)E^{-\Delta}\right]}.
\end{eqnarray}
It is worthy to mention that the value, $q \simeq 1/2$, can be obtained in the previous expression by considering the best fit values of the cosmological parameters only for high redshift, i.e., $H \gg H_{0}$. This implies a long matter dominated epoch for this kind of model and this is desired if we want to have formation of structures from primordial inhomogeneities; this behavior for the deceleration parameter is similar to the one obtained in \cite{waga}. As in the simple model case, the generalized approach exhibits a transition from decelerated to accelerated cosmic expansion. However, such transition takes place around the value $0.7$ for the redshift, then this scheme seems to be in more agreement with the value for the transition redshift reported in \cite{Moresco:2016mzx}, which was obtained without invoking a specific cosmological model. As can be seen in the figure, the parameter state for the holographic dark energy decreases as universe evolves and at present time is close to the value $-1$, such value for $\omega$ represents a cosmological constant behavior but due to matter creation effects, the model behaves as a quintessence fluid at effective level.

To end this section we consider the coincidence parameter, $r$, for this kind of cosmological model. The definition of the coincidence parameter is given as
\begin{equation}
    r := \frac{\rho_{m}}{\rho_{x}},
\end{equation}
then, using the Friedmann constraint (\ref{eq:fried1}) and (\ref{eq:roxhol1}) we obtain for the generalized model
\begin{equation}
    r(z) = \frac{E^{\Delta}}{\gamma}-1,
\end{equation}
if we evaluate at present time the previous expression we get, $r(z=0) = \gamma^{-1}-1 \approx 0.47$. It is worthy to mention that this value is close to the one obtained in the $\Lambda$CDM model which is approximately equal to $0.43$. On the other hand, if we solve for the redshift the condition $r(z)=1$ or equivalently $\rho_{x} = \rho_{m}$; we have $E^{\Delta} = 2\gamma$. Therefore the dominance of dark energy for this model started at the recent past, $z=0.363724$. For the $\Lambda$CDM it is estimated that began around $z \approx 0.55$. For the simple model we observe that the coincidence parameter takes a constant value
\begin{equation}
    r = \frac{1}{b^{2}}-1,
\end{equation}
this is expected since both densities, both $\rho_{m}$ and $\rho_{x}$ are proportional to $H^{2}$. Then the condition $\rho_{m} = \rho_{x}$ leads to $b=\sqrt{1/2}$, it is not possible to establish from the coincidence parameter the redshift at which the universe began to accelerate. However, such transition can be visualized in Fig. (\ref{fig:qdzsimp}). In fact, for the full model we observe a discrepancy between the values obtained for the beginning of the accelerated expansion by means of the deceleration and coincidence parameters. In this sense, we consider that the coincidence parameter for this kind of cosmological model it is not a good estimator for the beginning of the domination era of dark energy.

\subsection{Three special cases}

In this section we consider three special cases of our model. The first model (i) is the one with $\alpha=0$ in Eq. (\ref{eq:waga}) and represents a constant particle rate production that seems to be preferred by the previous analysis, (ii) the case with $\alpha=-1$ represents the well known case for a matter creation model that mimics the cosmological constant \cite{cardenas1}, and (iii) the case $\Delta=0$ which is a minimal extension of the simple model discussed previously.  
\begin{figure}[h!]
    \centering
    \includegraphics[width=8cm]{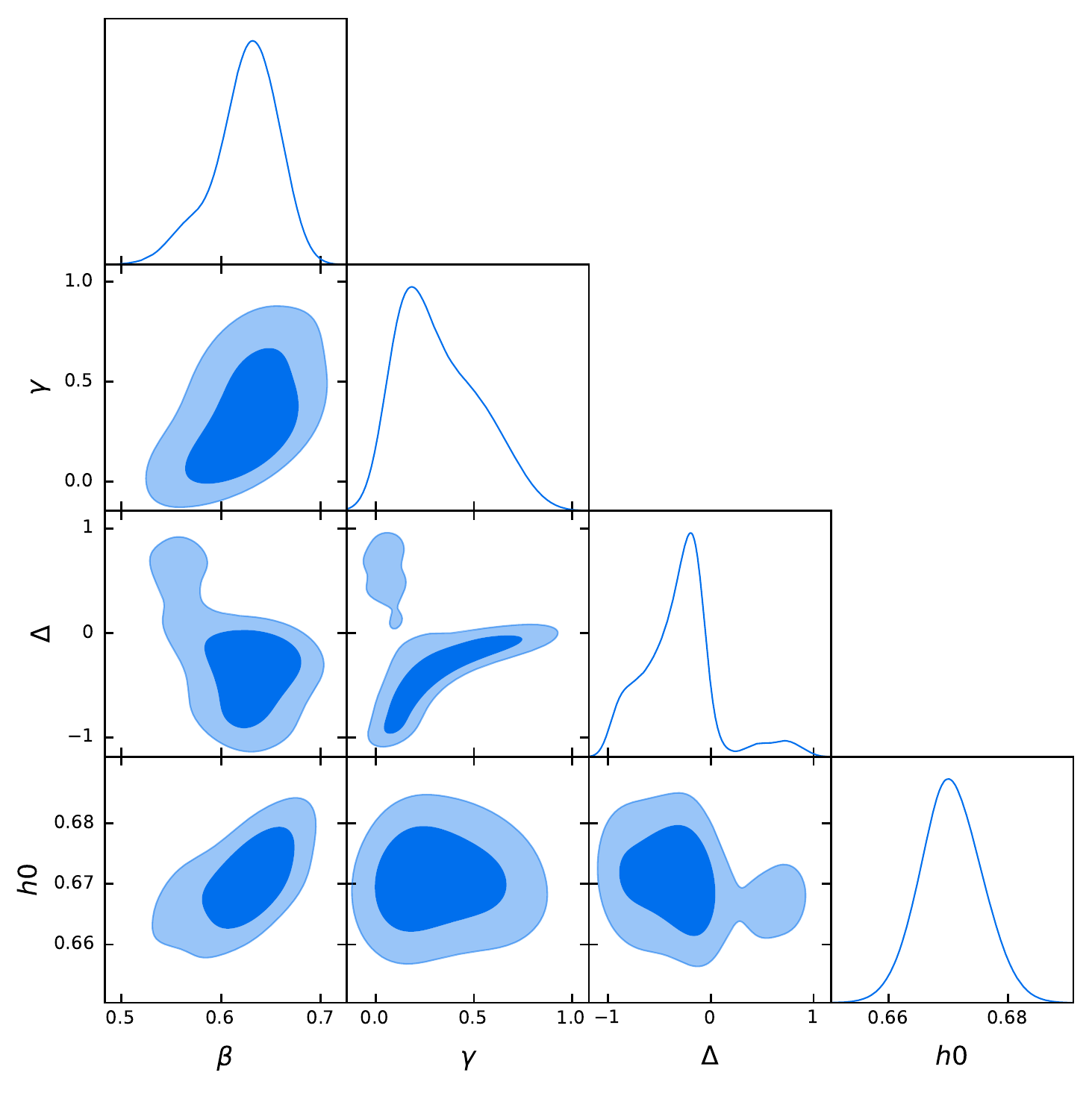}
    \caption{1D and 2D posteriors of the model with $\alpha=0$.}
    \label{fig:alfa0}
\end{figure}

The result of the analysis of the first case, with $\alpha=0$ is shown in Fig.(\ref{fig:alfa0}) and the best fit results are in the second column of Table (\ref{tab:results}). Notice that except for $h$ -- that is natural because we are using a prior on it -- all the others parameters have changed compared to the full model (first column in Table (\ref{tab:results})). Notice that the best fit value for $\Delta$ is consistent with zero, meaning that the holographic dark energy behaves as $\rho_x \propto H^2$. This implies that the accelerated expansion is driven completely by the matter creation process with $\beta \simeq 0.62$. Note that that this value for $\beta$ is distant from the value obtained for the full model.

Now we can understand the result in the full model. There, the matter creation process is not the source of the cosmic acceleration, in this case the accelerated expansion is driven by the holographic dark energy because $\beta \simeq 0$ but the Barrow exponent, $\Delta \simeq 1.5$, modifies the holographic dark energy evolution.
\begin{figure}[h!]
    \centering
    \includegraphics[width=8cm]{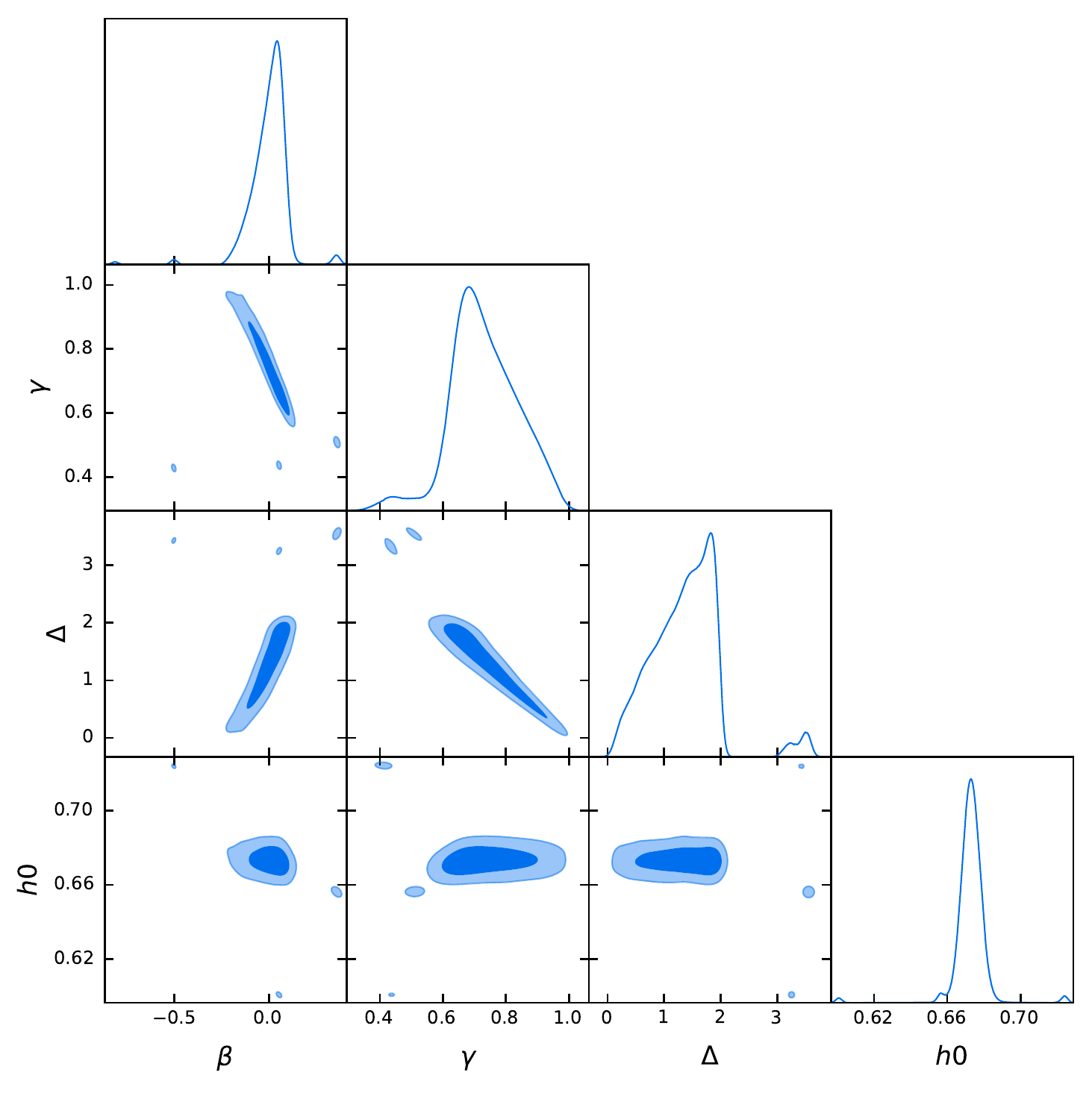}
    \caption{1D and 2D posteriors of the model with $\alpha=-1$.}
    \label{fig:alfa1}
\end{figure}

Next we discuss the case with $\alpha=-1$. In the case of a pure  matter creation model, i.e. in a universe with only DM and the matter creation process, it is well known that this model mimics exactly $\Lambda$CDM, being $\beta$ in (\ref{eq:waga}) the parameter that takes the role of the cosmological constant \cite{waga, cardenas1}. The result shown in Fig.(\ref{fig:alfa1}) and the best fit values displayed in the third column of Table (\ref{tab:results}), tell us that we have the opposite behaviour compared to the previous case. Here the matter creation process is not operating because the best fit value for $\beta$ is consistent with zero. Instead, as the case for the full model (see the first column of Table (\ref{tab:results})) the source of the accelerated expansion is the Barrow holographic term with $\Delta \simeq 1.4$.

\begin{figure}[h!]
    \centering
    \includegraphics[width=8cm]{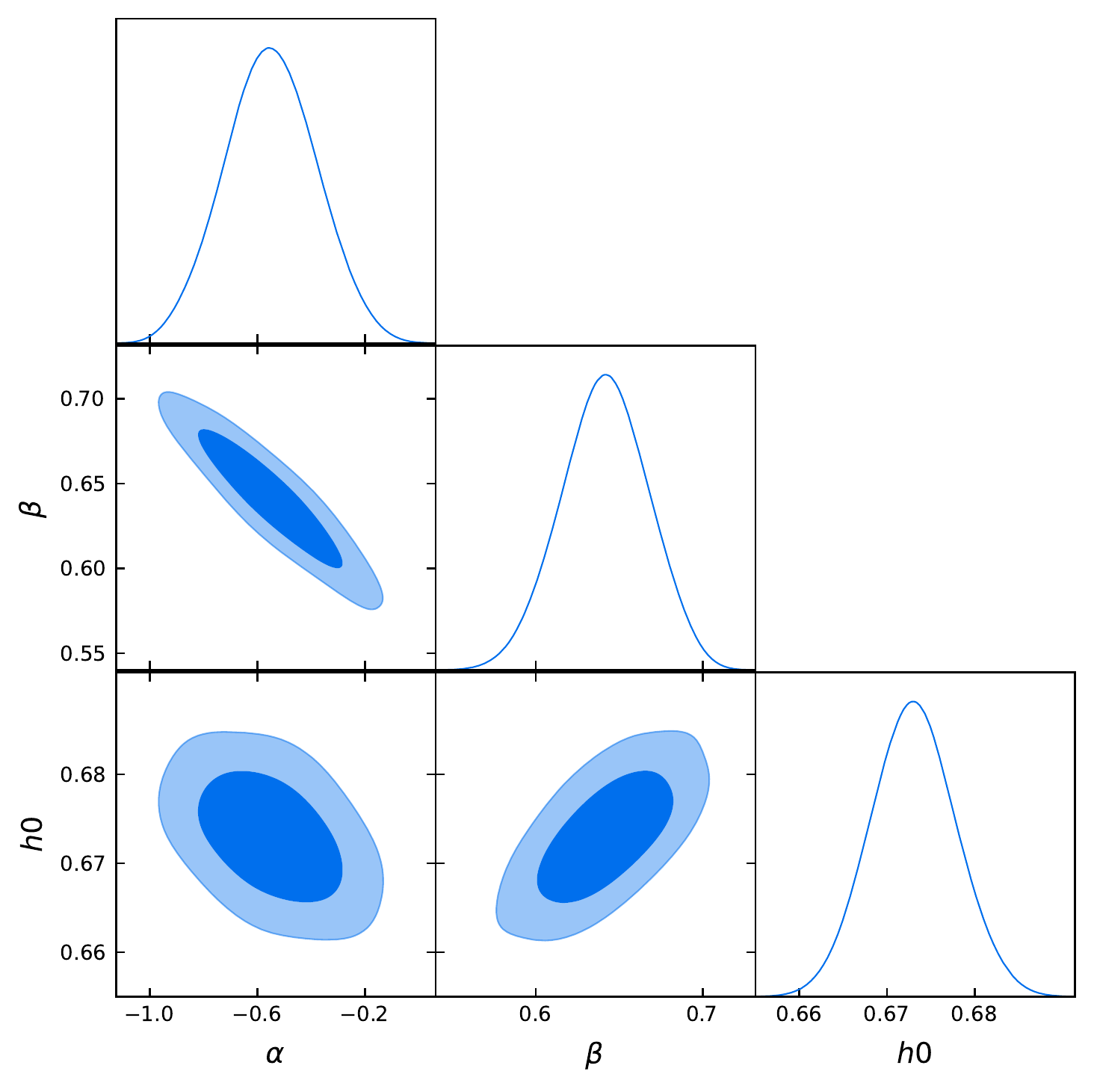}
    \caption{1D and 2D posteriors of the model with $\Delta = 0$.}
    \label{fig:Delta0}
\end{figure}

The last case is the one with $\Delta=0$. Here we are forcing the system to work with the Cohen version of the holographic DE, the one we know can not drive the acceleration of the cosmic expansion. This is a slight generalization of the simple model discussed in the previous section, now enabling to have a more complex particle creation rate function $\Gamma$. As we can check by setting $\Delta$ to zero in (\ref{eq:turning2}), the normalized Hubble function $E(z)$ does not depend on $\gamma$, and the quest for the best fit values can only constrain $\alpha, \beta$ and $h$. The result is shown in Fig.(\ref{fig:Delta0}) and the best fit values are displayed in the last column of Table (\ref{tab:results}). Because the source that drives the acceleration is the matter creation process, then is natural to expect a non zero $\beta \simeq 0.64$. The exponent, with the suggestive behaviour $\alpha \simeq -0.5$ seems to do a better job than setting $\alpha =0$ as done for the simple model discussed first.

Only for comparative reasons between our results and $\Lambda$CDM, we can write for the Barrow holographic dark energy, $\Omega_{x}(z) = \gamma E^{-\Delta}$, then at present time we have $\Omega_{x,0} = \gamma$. In the concordance model it is well established that the age of the oldest objects in the universe like stellar populations or globular clusters is helpful for setting a lower bound on $\Omega_{\Lambda}$. According to recent estimations for the age of these objects one obtains $\Omega_{\Lambda} \approx 0.6$ and for a successful structure formation such value must not be much larger than 0.7 \cite{velten}. Therefore, according to our results, the full model and the one labeled as case (ii) could lead to viable descriptions of our observable universe.     

\section{Conclusions and perspectives}
\label{sec:final}

In this work we have explored a cosmological proposal that merges two interesting scenarios: matter creation effects and holographic dark energy. This idea was studied before in Refs. \cite{sing, singb}. However, we consider that some of the authors' assumptions in those works are not physically consistent and require scrutiny. We must mention that our approach is an extension of the one studied in \cite{sing, singb} and consists in a generalization for the holographic cutoff given by the introduction of the Barrow exponent ($\Delta$) and a variable particle production rate characterized by the parameters $\alpha, \beta$. The values $\Delta = \alpha = 0$ in our approach,  lead to the simple model explored in the aforementioned references. As stated in the work, we restrict ourselves to the use of the Hubble scale as characteristic length for the holographic dark energy since we are evaluating the contribution coming from matter creation effects to the late times accelerated cosmic expansion. Our proposal was tested against observations without any a priori assumption for the free cosmological parameters, the results obtained from the statistical analysis are summarized in Table (\ref{tab:results}). For the full contribution of both effects, astrophysical observations lead to a scenario where the value of the Barrow exponent greatly modifies the holographic cutoff and in such case the cosmic expansion is mostly driven by the holographic dark energy. For a constant particle production rate, dubbed as case (i), we have that the cosmic acceleration has its origin in matter creation effects. On the other hand, case (ii), which corresponds to a resembling of the $\Lambda$CDM model within the matter creation scheme, leads again to a scenario where the holographic dark energy plays a relevant role driving the cosmic expansion. Finally for case (iii) the Barrow exponent is turned off, observations indicate that matter productions effects are enough to drive the cosmic expansion since the values obtained for the parameters $\alpha$ and $\beta$ allow to maintain the generalized form of the rate function, $\Gamma$.

At effective level, if we consider the best fit values obtained for the cosmological parameters, the parameter state for the total energy density in general evolves as a function of the redshift and at present time we obtain a quintessence behavior in the model if we compare it with the single fluid description. On the other hand, for the holographic dark energy we also observe a dynamical behavior of its parameter state and closeness with the $\Lambda$CDM model at present time. The transition observed in the effective parameter state from a decelerated stage of the cosmic expansion to an accelerated one is confirmed by the deceleration parameter computed from the model. At some redshift value we observe that such parameter becomes negative. It is worthy to mention that for this kind of cosmological model the coincidence parameter does not provide clear information about the beginning of dark energy dominance, for instance, in the simple model the deceleration parameter, indicates a transition from deceleration to acceleration whilst the coincidence parameter takes a constant value, i.e, the ratio between dark matter and dark energy densities does not change. Despite the consistent results of the model, a missing scenario in this cosmological approach is the phantom regime. This regime is not discarded at all by observations and some adjustments in the model could provide a crossing to the phantom zone at effective level; in Ref. \cite{usaj} a first attempt in this sense can be found but only under the consideration of matter creation effects. Phantom cosmology deserves further investigation since it is well known that the thermodynamics description of this regime has some issues, see \cite{us} and references therein. We leave this subject open for future investigation.  

\section*{Acknowledgments}
M.C. work has been supported by S.N.I. (CONACyT-M\'exico).

\end{document}